\documentclass[prb,aps,twocolumn,superscriptaddress,showpacs,footinbib]{revtex4-2}
\usepackage[colorlinks=true,linkcolor=black, citecolor=blue, urlcolor=blue,
    unicode=true]{hyperref}

\usepackage{graphicx,amsmath,braket,tikz,amssymb,footmisc,lipsum,array,tabularx,float,comment}

\usepackage{cleveref}

\usepackage[utf8]{inputenc}
\usepackage{color}

\usepackage{subfigure}

\usepackage{multirow}

\crefname{equation}{}{}
\Crefname{equation}{}{}
\crefrangelabelformat{equation}{#3#1#4--#5#2#6}

\crefmultiformat{equation}{#2#1#3}{, #2#1#3}{#2#1#3}{#2#1#3}
\Crefmultiformat{equation}{#2#1#3}{, #2#1#3}{#2#1#3}{#2#1#3}

\allowdisplaybreaks

\newcommand{\cso}{CsO$_{2}$ }
\newcommand{\rbo}{RbO$_{2}$ }

\begin{document}


\title{The crystal and magnetic structure of cesium superoxide}

\author{R. A. Ewings}
\email[]{russell.ewings@stfc.ac.uk}
\affiliation{ISIS Pulsed Neutron and Muon Source, STFC Rutherford Appleton Laboratory, Harwell Campus, Didcot, Oxon, OX11 0QX, United Kingdom}

\author{M. Reehuis}
\affiliation{Helmholtz-Center Berlin for Materials and Energy, D-14109 Berlin, Germany}

\author{F. Orlandi}
\affiliation{ISIS Pulsed Neutron and Muon Source, STFC Rutherford Appleton Laboratory, Harwell Campus, Didcot, Oxon, OX11 0QX, United Kingdom}

\author{P. Manuel}
\affiliation{ISIS Pulsed Neutron and Muon Source, STFC Rutherford Appleton Laboratory, Harwell Campus, Didcot, Oxon, OX11 0QX, United Kingdom}

\author{D. D. Khalyavin}
\affiliation{ISIS Pulsed Neutron and Muon Source, STFC Rutherford Appleton Laboratory, Harwell Campus, Didcot, Oxon, OX11 0QX, United Kingdom}

\author{A. S. Gibbs}
\affiliation{ISIS Pulsed Neutron and Muon Source, STFC Rutherford Appleton Laboratory, Harwell Campus, Didcot, Oxon, OX11 0QX, United Kingdom}
\affiliation{School of Chemistry, University of St Andrews, North Haugh, St Andrews, United Kingdom}

\author{A. D. Fortes}
\affiliation{ISIS Pulsed Neutron and Muon Source, STFC Rutherford Appleton Laboratory, Harwell Campus, Didcot, Oxon, OX11 0QX, United Kingdom}

\author{A. Hoser}
\affiliation{Helmholtz-Center Berlin for Materials and Energy, D-14109 Berlin, Germany}

\author{A. J. Princep}
\affiliation{ISIS Pulsed Neutron and Muon Source, STFC Rutherford Appleton Laboratory, Harwell Campus, Didcot, Oxon, OX11 0QX, United Kingdom}
\affiliation{Department of Physics, University of Oxford, Clarendon Laboratory, Oxford OX1 3PU, United Kingdom}

\author{M. Jansen}
\affiliation{Max-Planck-Institut f\"{u}r Festk\"{o}rperforschung, D-70569 Stuttgart, Heisenbergstr. 1, Germany}

\date{\today}

\begin{abstract}

\cso is a member of the family of alkali superoxides (formula $A$O$_{2}$ with $A=$ Na, K, Rb and Cs) that exhibit magnetic behavior arising from open $p$-shell electrons residing on O$_{2}^{-}$ molecules. We use neutron diffraction to solve the crystal and magnetic structures of \cso, and observe a complex series of structures on cooling from room temperature to $1.6$\,K. These include an incommensurate modulation along the $a$-axis of the structure at intermediate temperatures, which then locks into a commensurate modulation that doubles the unit cell compared to the previously supposed orthorhombic unit cell. In both incommensurate and commensurate phases our structural solution involves a staggering of the cesium ion positions along the $b$-axis, in contrast to studies of other alkali superoxides in which staggered tilts of the O$_{2}^{-}$ dimers relative to the $c$-axis are seen. Below $T\simeq 10$\,K we observe magnetic Bragg reflections arising from an antiferromagnetically ordered structure with a wavevector of $\mathbf{k}=(0,0,0)$ (relative to the doubled crystallographic unit cell), with moments that point predominantly along the $b$-axis with a small component along the $a$-axis that hints at possible anisotropic exchange coupling (consistent with the crystal structure). Measurements of the magnetic Bragg reflections in an applied magnetic field suggest a spin-flop transition takes place between 2\,T and 4\,T in which moments likely flop to point along the crystallographic $a$-axis. Our measurements indicate that CsO$_{2}$ is an interesting example of magnetic properties being inherently linked to the crystal structure, in that the staggered displacement of the cesium ions activates antisymmetric exchange which then permits the observed spin canting.

\end{abstract}

\maketitle

\section{Introduction}\label{sec:intro}

The complex interplay between spin, orbital and lattice degrees of freedom plays a varied role across many material families, resulting in a plethora of emergent properties. Well known examples include the Jahn-Teller interaction in transition metal oxides \cite{BlundellMag}, magnetoelastic coupling \cite{CallenCallen_magnetoelastic}, and quantum spin liquid states in strongly spin-orbit coupled materials \cite{Rau_QSL_review}. This latter example is particularly topical, with the search ongoing for a material showing the bond-directional Ising-like anisotropy necessary for a realisation of the Kitaev model, for instance.

Anionogenic magnetic materials are compounds on which the spin degree of freedom is associated with the partially occupied $p$-orbitals of an anionic molecule. Although theoretical predictions exist for other simple anions \cite{Volnianska2010}, the most well-known experimental examples are for the cases of oxygen anions. Some such materials known to be magnetic include solid oxygen \cite{Stephens1986,Felix2008,Klotz2010}, alkali sesquioxides, and alkali superoxides \cite{Hesse1989}. In the latter, the magnetic moment arises due to a partially (half) occupied $\pi^{\star}$ orbital on the O$_{2}^{-}$ molecular units. Because the magnetic object is spatially extended, on so-called O$_{2}^{-}$ dumbbells, rather than arising from unpaired electrons on a single ion, a natural coupling arises between the magnetism and the crystal structure. Of particular interest is the interplay of magnetism with the orientation of the dumbbells, a possibility which was noted explicitly in the 1980s and referred to as magnetogyration \cite{Boesch1980}. The alkali sesquioxides have attracted much recent attention because they undergo charge ordering which then leads to the formation of magnetic pyrochlore or dimer lattices \cite{Winterlik2009,Arcon2013,Adler2018,Colman2019,KnaflicPRB2020}.

The alkali superoxides $A$O$_{2}$ ($A$ = Na, K, Rb, Cs) have been known for a relatively long time, with studies in the 1970s and 1980s elucidating many of the key physical properties. In particular, the fact that all of them are magnetic and undergo a cascade of structural phase transitions on cooling was recognized at this time \cite{Hesse1989}. The materials in the family have a common high temperature structure of the NaCl type \cite{Ziegler1976} in which the O$_{2}^{-}$ dumbbells are able to rotate freely. As they are cooled the rotation of the dumbbells progressively becomes hindered, so they precess about the $c$-axis of a tetragonal structure, and then eventually freeze. The details of this freezing are fairly well characterized for KO$_{2}$, which forms a monoclinic structure at low temperature in which the dumbbells are frozen at an angle from the higher temperature tetragonal $c$-axis \cite{Halverson1962,Smith1966}. Neutron diffraction measurements from the 1960s confirm that KO$_{2}$ orders antiferromagnetically \cite{Smith1966}, however in the data collected at that time only two magnetic Bragg peaks could be discerned above background, making a magnetic structure refinement impossible. On the other hand, the precise details of the structures of \rbo and \cso were much less well characterized at that time \cite{Ziegler1976,Rosenfeld1978}. What was known is that, like KO$_{2}$, both materials take on a tetragonal structure at room temperature, with transitions at $\sim 150$\,K into possible orthorhombic structures. Around the same temperature an incommensurate structure was observed in both materials, with a period around three lattice units along the $a$-axis. Neither the orthorhombic structure nor the incommensurate structure were solved in detail in either case, however.

More recent experimental studies of CsO$_{2}$, RbO$_{2}$ and NaO$_{2}$ have revisited their bulk magnetic properties. In particular, \cso has been posited to be a pseudo-1-dimensional magnet, with evidence including a broad hump in the magnetic susceptibility reminiscent of what is predicted by a modified Bonner-Fisher model \cite{Bonner1965}. Magnetic order occurs at $\sim 10$\,K, and DFT calculations of superexchange pathways \cite{Riyadi2012}, and NMR \cite{Klanjsek2015} and EPR \cite{Knaflic2015} data are consistent with the formation of a Tomonaga-Luttinger liquid phase.

These studies have also revisited the crystal structure, noting the presence of a transition from a tetragonal to an orthorhombic structure below $\sim 70$\,K (a substantially lower temperature than was seen for a similar transition in earlier studies). Notwithstanding, the crystallographic details of the orthorhombic structure were not solved, and nor were the details of the magnetic structure. High field magnetization studies \cite{Miyajima2018} are consistent with \cso exhibiting reduced dimensionality, however these data are not fully consistent with the models proposed in the other studies, and the authors note that in the absence of detailed knowledge of the magnetic structure all of the data are hard to interpret. Recent studies combining $\mu$SR, Raman spectroscopy, and x-ray diffraction on \rbo \cite{Astuti2019} and NaO$_{2}$ \cite{Miyajima2021} have been interpreted similarly, with progressively lowering symmetry on cooling, including the observation of incommensurate structural Bragg peaks in RbO$_{2}$. The main difference between the two compounds is that the data on \rbo are more consistent with a 3-dimensional Heisenberg model rather than 1-dimensional magnetism, whereas the data on NaO$_{2}$ suggest that there is low dimensionality and no long range magnetic order.

Application of a relatively modest magnetic field, of $\sim 2.5$\,T at 2\,K, results in a significant change in the magnetic susceptibility of \cso \cite{Miyajima2018}. This is interpreted as being a spin flop transition from a low field antiferromagnetic state with easy-axis anisotropy. It is notable that the majority of the NMR measurements interpreted as evidence for the formation of a Tomonaga Luttinger liquid were taken at fields either close to or above the spin flop transition field.

As already discussed, the crystal and magnetic structure of KO$_{2}$ is better known, and perhaps as a consequence of this there has been more theoretical interest in this member of the superoxide family. Studies have focused on the coupling between spin, orbital and lattice physics. Several authors have proposed that KO$_{2}$ undergoes orbital ordering \cite{Nandy2010,Kim2010,Kim2014}, which then has consequences for the nature of superexchange between O$_{2}^{-}$ molecules via the K ion and hence affects the magnetic order that develops. In particular, anisotropic exchange has been discussed \cite{Solovyev2008} which as well as affecting the propensity to magnetic order may also give rise to unusual magnetic excitations. The calculations rely on a good knowledge of the crystal structure to have predictive power, underscoring the importance of accurate measurements.

Given the ongoing ambiguity concerning the low temperature crystal structure of \cso, and the ordered magnetic structure, we were motivated to revisit this problem using modern neutron scattering instrumentation and analysis methods. As well as providing a complete picture, such measurements are invaluable for the interpretation of data from other experimental probes as well as informing future theoretical predictions. We found that at low temperature the crystallographic unit cell is doubled along the $a$-axis compared to the room temperature tetragonal cell, with a staggered displacement along the $b$-axis of the Cs ions but essentially no tilts of the O$_{2}^{-}$ dumbbells. This phase sets in below 80\,K. We also observed the incommensurate crystal structure, akin to one seen in early studies, and found that like the low temperature phase it is characterized by an $a$-axis modulation of the positions of the Cs ions along the $b$-axis. It would seem that this incommensurate structure eventually locks into the aforementioned commensurate one on cooling. The incommensurate phase is visible below 190\,K, below which temperature there is a distinct transition from the room temperature tetragonal phase (space group $I4/mmm$) to another that, aside from the incommensurate reflections, can be indexed using $Immm$. In the magnetically ordered phase below 10\,K we found that the moments align antiferromagnetically, with the largest component along the $b$-axis but a small component along the $a$-axis also, indicating possible exchange anisotropy.

\section{Methods}\label{sec:methods}

The \cso powder samples used for this study were prepared using the well-established method of oxidation of distilled Cs metal with dried molecular oxygen \cite{Helms1939}.

Neutron diffraction experiments were conducted using the E2 \cite{E2ref}, E6 and E9 \cite{E9ref} diffractometers at the BER-II reactor at the Helmholtz Zentrum Berlin, and on the WISH \cite{WISH-tech} and HRPD \cite{HRPD-tech} diffractometers at the ISIS spallation neutron source. Experiments on E2 we performed using a fixed incident neutron wavelength of $\lambda = 2.38$\,\AA, selected by a PG monochromator, between temperatures of 1.7 and 100\,K in applied magnetic fields between 0 and 6.5\,T.

The sample was pressed into cylindrical pellets of diameter 6\,mm and height 3\,mm, ten of which were placed into a sealed quartz ampoule in which the sample was well fixed in order to avoid movement of the powder grains in an applied magnetic field. This ampoule was placed into a cylindrical vanadium container of diameter 8\,mm and height 60\,mm. Experiments on E6 were performed with a fixed incident neutron wavelength $\lambda = 2.41$\,\AA, also selected by a PG monochromator, for temperatures between 1.6 and 248\,K in zero applied magnetic field, using the same sample containment as for E2. One dataset was collected using E9, with a fixed incident neutron wavelength of $\lambda = 1.7985$\,\AA, selected by a Ge monochromator, at a temperature of 4\,K, again with the same sample containment. Neutron powder patterns were collected between the diffraction angles 7.8$^{\circ}$ to 83.4$^{\circ}$ (E2), 5.5$^{\circ}$ to 136.4$^{\circ}$ (E6), and 7.5$^{\circ}$ to 141.7$^{\circ}$ (E9), respectively.

HRPD and WISH are both time-of-flight (ToF) instruments, with a white beam of neutrons incident on the sample and analysis of the arrival time at the detectors (ToF) used to determine wavelength and hence Bragg reflection d-spacing. Experiments on HRPD were performed at temperatures of 20\,K and 80\,K, with further measurements taken over fixed time intervals (6 minutes) as the sample was cooled from 290\,K. The sample was loaded into an 8\,mm diameter vanadium can of height 40\,mm for these measurements. All of the data used for subsequent analysis were collected in the highest angle detector bank, with data time-focused to a scattering angle of $2\theta = 168.33^{\circ}$. Experiments on WISH were performed at temperatures in the range 1.6 to 300\,K. The sample was loaded into an aluminum can with an annular geometry, with height 40\,mm, outer diameter 20\,mm and sample thickness 2\,mm. The sample was first cooled to 1.6\,K, then a measurement of 40 minutes duration was performed at 1.6\,K, followed by another of the same duration at 25\,K. The sample was then re-cooled to 1.6\,K and further measurements were made at fixed temperatures for 5 minutes per temperature, warming back up to 300\,K. A sample temperature fluctuation of $\leq 2\%$ was observed for each of the measurements. For crystal structure analysis data time-focused on to the higher angle detector banks, at $2\theta = 152.9^{\circ}$ and $2\theta = 121.7^{\circ}$ were used. For the magnetic refinements the lower angle detector bank data, at $2\theta = 58.3^{\circ}$ was used.

Structural refinements were performed using the FullProf software suite \cite{FullProfStruc,FullProfMag} for the E2, E6 and E9 data, and using Jana2006 \cite{JanaSoftware} for the HRPD and WISH structural data. Subsequent magnetic refinements were performed using FullProf for the WISH data. Magnetic symmetry analysis for the magnetic structure refinement was done using the Basirreps tool implemented in FullProf. The incommensurate structure analysis was done using Jana2006 together with the Isodistort software \cite{Isotropy-url,Isotropy-article}.

\section{Results}\label{sec:results}

\subsection{Crystal structure}
\begin{figure}[!h]
\centering
    \includegraphics[scale=0.6]{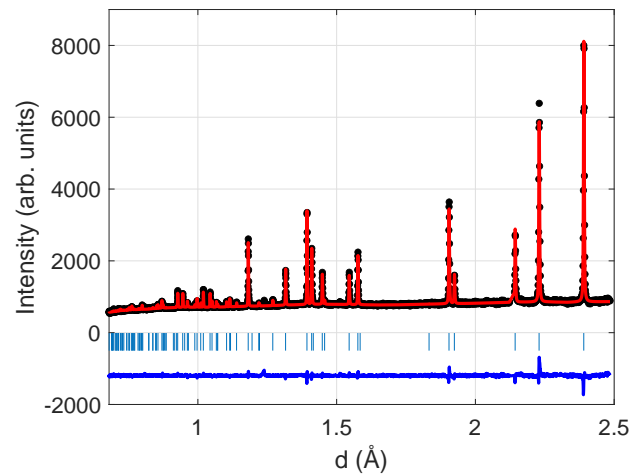}
    \caption{Data collected at 290\,K on HRPD as a function of d-spacing, refined using the $I4/mmm$ space group (see text). Measured data are indicated by black circles, the fit by the red line, and the difference curve (offset by $-1200$ units) is shown as a blue curve. Vertical ticks indicate the positions of Bragg reflections. The residuals for these fits were $R_{p}=2.18$ and $R_{wp}=2.88$.}
    \label{f:HRPD290}
\end{figure}

Figure \ref{f:HRPD290} shows data collected at 290\,K (room temperature) on the HRPD instrument, together with a refinement with the $I4/mmm$ space group and lattice parameters widely reported in the literature. A satisfactory refinement can be achieved for these data collected for a relatively short duration (6 minutes) with correspondingly non-optimized signal to noise.

The details of the crystal structure determined at 290\,K are given in table \ref{tab:I4mmm_struc}. Refinements of the data from WISH at the same temperature are consistent with these parameters.

\begin{table}
\caption{Crystal structure parameters for the $I4/mmm$ space group solution, at 290\,K measured on the HRPD instrument. The determined lattice parameters were $a = 4.4589(2) \AA$ and $c = 7.3320(3)\AA$. The unit cell volume was $V = 145.796(1) \AA^{3}$. The residuals for the fit were $R_{p}=2.18\%$ and $R_{wp}=2.88\%$.}
\centering
\begin{tabular}{ p{10mm}p{8mm}p{15mm}p{15mm}p{15mm}p{15mm}  }
 \hline \hline
 \vspace{1mm} & \vspace{1mm} & \vspace{1mm} & \vspace{1mm} & \vspace{1mm} & \vspace{1mm} \\
 Atom & Site & $x$ & $y$ & $z$ & $U_{\rm{iso}} (\AA^{2})$\\
 \vspace{1mm} & \vspace{1mm} & \vspace{1mm} & \vspace{1mm} & \vspace{1mm} & \vspace{1mm} \\
 \hline
 \vspace{1mm} & \vspace{1mm} & \vspace{1mm} & \vspace{1mm} & \vspace{1mm} & \vspace{1mm} \\
 Cs & $2a$ & 0.00 & 0.00 & 0.00 & 0.0323(6)\\
 O & $4e$ & 0.00 & 0.00 & 0.4138(1) & 0.0404(5))\\
 \vspace{1mm} & \vspace{1mm} & \vspace{1mm} & \vspace{1mm} & \vspace{1mm} & \vspace{1mm} \\
 \hline
\end{tabular}
\label{tab:I4mmm_struc}
\end{table}

\begin{figure}[!h]
\centering
    \includegraphics[scale=0.44]{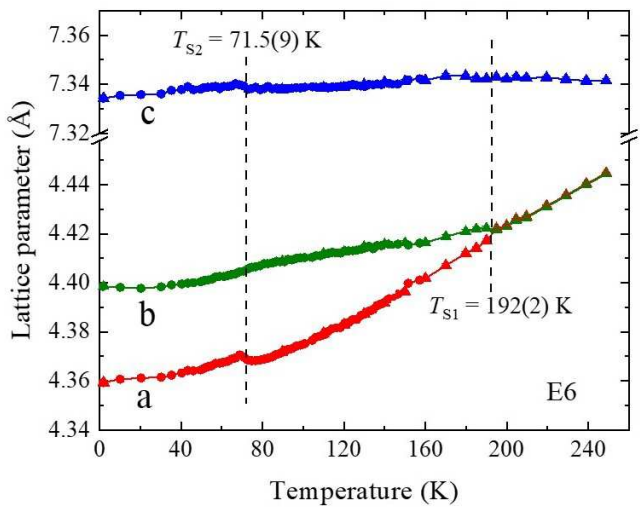}
    \caption{Temperature dependence of lattice parameters refined from E6 data, assuming $I4/mmm$ and $Immm$ space groups at high and low temperature respectively.}
    \label{f:Tdep_latt}
\end{figure}

On cooling a phase transition is observed at $T_{\rm{S1}} = 192(2)$\,K, that appears to be tetragonal to orthorhombic as illustrated in fig. \ref{f:Tdep_latt} which shows the lattice parameter as a function of temperature determined from Rietveld refinements using the tetragonal $I4/mmm$ space group above the transition and the orthorhombic $Immm$ space group below it, from data collected on E6. The latter space group represents the symmetry of the lattice if only macroscopic strain is taken into account \cite{Isotropy-article,Isotropy-url}. The strain reflects the change of the unit cell metric below the transition and therefore this space group is sufficient to evaluate the thermal evolution of the unit cell parameters. The transition is at a temperature consistent with early structural studies of \cso \cite{Ziegler1976}, but significantly higher than observed in more recent x-ray diffraction experiments \cite{Riyadi2012}. In order to get good fits we found that anisotropic displacement parameters (ADPs) were required. The best fits were obtained with considerable elongation of the ADP ellipsoids of the cesium ions along the $b$-axis compared to the other principal axes. Furthermore, the ADP ellipsoids of the oxygen ions were elongated in the $ab$-plane, albeit with a smaller difference between $a$ and $b$ than for the cesium ions, and compressed significantly along the $c$-axis. These findings suggest the presence of atomic displacements not accounted for in the $Immm$ space group.

\begin{figure}[!h]
\centering
    \includegraphics[scale=0.58]{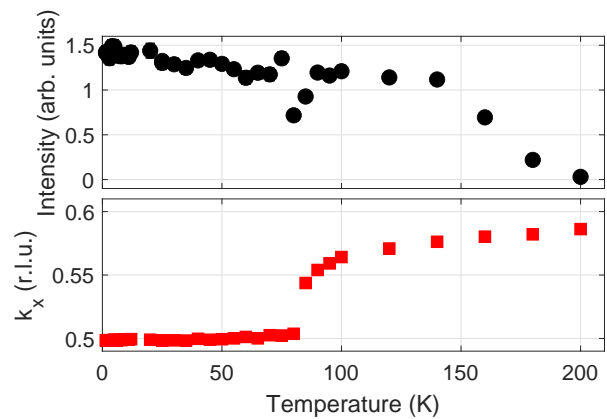}
    \caption{Data from WISH showing the disappearance on warming up to 200\,K of the incommensurate Bragg reflection at $\mathbf{Q}_{\rm{IC}} = (1.56,2,1)$ at $d = 1.69$\,\AA. The upper panel shows the intensity of the peak, determined by integrating a fitted Gaussian, with black circles. The lower panel shows the fitted position of the peak $k_{x}$ with respect to $(1 + k_{x},2,1)$ with red squares. In both cases the errorbars are smaller than the point size.}
    \label{f:ic_tdep}
\end{figure}

Close examination of the data collected on WISH shows that at temperatures above $\sim 72$\,K, and at least up to 180\,K, weak incommensurate superlattice reflections are visible above background that can be indexed using a propagation vector of $\mathbf{k}_{\rm{IC}} = (0.561(2),0,0)$ referred to the tetragonal unit cell, i.e. a period of 1.78 lattice units. The appearance of the incommensurate superlattice seems to be concomitant with the putative transition from the tetragonal to orthorhombic structure at $T_{\rm{S1}}$. To probe this further we show in fig. \ref{f:ic_tdep} the temperature dependence of the intensity and position of the $(1 + k_{x},2,1)$ peak (the fact that superlattice peaks such as this one appear at short $d$-spacing, and hence large $|\mathbf{Q}|$, indicates that they are structural rather than magnetic in origin).

\begin{figure}[!h]
\centering
    \includegraphics[scale=0.33]{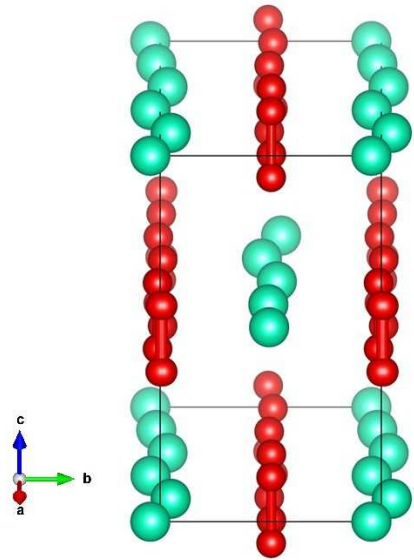}
    \caption{Representation of the incommensurate crystal structure determined using the refined WISH data at 100\,K. The Cs ions (green spheres) and oxygen dumbbells (red spheres connected by red cylinder to indicate the molecular bond) modulate sinusoidally along the $a$-axis, with displacements along the $b$-axis (in tetragonal notation, which is cyclically permuted with respect to the notation in table \ref{tab:IC_struc}). Five unit cells along $a$ are shown in order to illustrate the form of the displacements. This image, and others in the rest of the manuscript illustrating crystal and magnetic sturctures, were rendered using the VESTA software \cite{MommaVESTA}.}
    \label{f:ic_struc}
\end{figure}

Given that the incommensurate reflections seem to appear at the same temperature as the putative tetragonal to orthorhombic transition, to solve the structural modulation we assumed that the corresponding order parameter is transformed by a single irreducible representation (irrep) of the high temperature tetragonal space group $I4/mmm$. The irreducible nature of the order parameter was concluded based on the continuous character of the transition evidenced by the temperature dependence of the unit cell parameters (fig 2). There are four irreps, $\Sigma_i (i=1-4)$ associated with the $(g,0,0)$ line of symmetry. The relevant superspace groups were generated by ISODISTORT \cite{Isotropy-article,Isotropy-url} and compared with the available diffraction data. Some of them could be ruled out due to extinction conditions that predict no intensity where a finite signal was observed (e.g. at $d = 4.05 \AA$). The remainder were tested using the Jana2006 software. The best agreement (with $R_{wp} = 8.83\%$) was found for the model with $Immm(0,0,g)s00$ symmetry yielded by the $\Sigma_4$ irrep and illustrated in fig \ref{f:ic_struc}. The largest displacements are of the cesium ions, which are staggered along the $b$-axis following a sinusoidal modulation with an amplitude of 0.0564(9) lattice units. The oxygen molecules also exhibit a sinusoidal displacement along $b$, albeit with a smaller amplitude of 0.0090(7) lattice units. Interestingly, this symmetry does not permit the rotation of oxygen molecules widely predicted for CsO$_{2}$, and seen in other alkali superoxides. Rather, the entire molecules’ centres of mass are displaced along $b$, with the molecules remaining oriented parallel to the c-axis. The complete structural solution for the incommensurate phase is given in table \ref{tab:IC_struc}.

\begin{table}
\caption{Structural parameters of CsO$_{2}$ obtained from refinement of the neutron diffraction data ($T = 100$\,K) using the $Immm(0,0,g)s00$ superspace group with the basis vectors related to the parent tetragonal $I4/mmm$ structure as $(0,-1,0,0),(0,0,1,0),(-1,0,0,0),(0,0,0,1)$ with the origin at $(0,0,0,\frac{3}{4})$. Here, $A_{i}^{1}$ and $B_{i}^{1}$ with $i=(x,y,z)$ are the Fourier coefficients of the first harmonic $(n = 1)$ of the displacive modulation function: $u_{i,j,l}(r_{j,l}, \mathbf{k}_{\rm{IC}})= \sum_{n=0}^{\infty} A_{i,j}^{n} sin (2 \pi n[\mathbf{r}_{j,l} \cdot  \mathbf{k}_{\rm{IC}}]) + B_{i,j}^{n} cos (2 \pi n[\mathbf{r}_{j,l} \cdot \mathbf{k}_{\rm{IC}}])$, where $\mathbf{r}_{j,l}$ indicates the position of the $j^{\rm{th}}$ atom of the average structure in the $l^{\rm{th}}$ unit cell. The unit cell parameters were $a = 4.4117(1) \AA$, $b = 7.3438(2) \AA$, $c = 4.3762(1)\AA$ and $\mathbf{k}_{\rm{IC}} = (0,0,0.561(2))$, the latter adopted for the setting of the superspace group specified above. The unit cell volume was $V = 141.782(6) \AA^{3}$. The residuals for the fit were $R_{p}=6.86\%$ and $R_{wp}=8.83\%$.}
\centering
\begin{tabular}{ p{15mm}p{17mm}p{17mm}p{17mm}p{12mm}  }
 \hline \hline
 \vspace{1mm} & \vspace{1mm} & \vspace{1mm} & \vspace{1mm} & \vspace{1mm} \\
 Atom & $x$ & $y$ & $z$ & $U_{\rm{iso}} (\AA^{2})$\\
 $A_{i}^{1}$ & $A_{x}^{1}$ & $A_{y}^{1}$ & $A_{z}^{1}$ & \\
 $B_{i}^{1}$ & $B_{x}^{1}$ & $B_{y}^{1}$ & $B_{z}^{1}$ & \\
 \vspace{1mm} & \vspace{1mm} & \vspace{1mm} & \vspace{1mm} & \vspace{1mm} \\
 \hline
 \vspace{1mm} & \vspace{1mm} & \vspace{1mm} & \vspace{1mm} & \vspace{1mm} \\
 Cs & 0.0 & 0.0 & 0.0 & 0.0029(5)\\
    & -0.0564(9) & 0.0 & 0.0 & \\
    & 0.0 & 0.0 & 0.0 & \\
 O & 0.0 & 0.40964(12) & 0.0 & 0.0110(3)\\
  & -0.0090(7) & 0.0 & 0.0 & \\
  & 0.0 & 0.0 & 0.0 & \\
 \vspace{1mm} & \vspace{1mm} & \vspace{1mm} & \vspace{1mm} & \vspace{1mm} \\
 \hline
\end{tabular}
\label{tab:IC_struc}
\end{table}

\begin{figure}[!h]
\centering
    \includegraphics[scale=0.58]{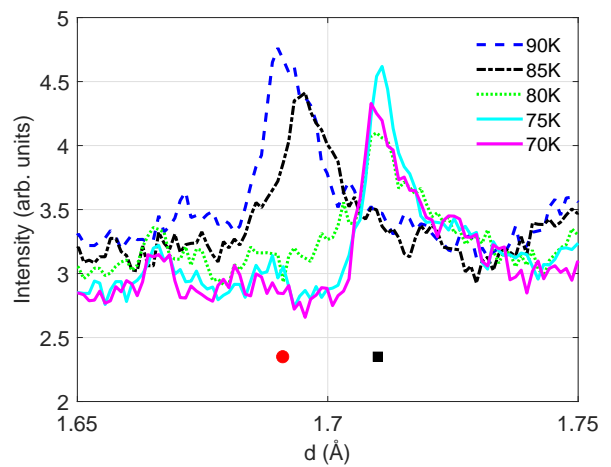}
    \caption{Data from WISH showing the transition from incommensurate to commensurate crystal structure. Below $\sim 80$\,K a peak corresponding to $\mathbf{Q}_{\rm{C}} = (\frac{3}{2},2,1)$ appears at $d = 1.71$\,\AA. At higher temperatures a peak corresponding to $\mathbf{Q}_{\rm{IC}} = (1.57,2,1)$ at $d = 1.69$\,\AA is visible. Commensurate and incommensurate peak positions are indicated by a black square and a red circle respectively.}
    \label{f:ic_to_c2}
\end{figure}

On cooling further a significant anomaly in the $a$ lattice parameter (in the $Immm$ setting), and a smaller anomaly in the $c$ parameter, occurs at $T_{\rm{S2}} = 71.5(9)$\,K. At the same time the propagation vector of the incommensurate structural modulation changes and locks in to a commensurate value of $\mathbf{k}_{\rm{C}} = (\frac{1}{2},0,0)$ (again referred to the $Immm$ cell), which survives down to the lowest temperatures measured. Data showing the transition locking into $\mathbf{Q}_{\rm{C}} = (\frac{3}{2},2,1) = (1,2,1) + \mathbf{k}_{\rm{C}}$ are shown in fig. \ref{f:ic_to_c2}.

In order to understand what is happening crystallographically in this phase we performed careful Rietveld refinements of all of the data collected on the different instruments used. In previous work it was suggested that the orthorhombic space group $Immm$ provides an appropriate description of the crystal structure. However, that would be inconsistent with the doubled unit cell seen below $T_{\rm{S2}}$.

\begin{table*}[ht]
\caption{Matrices of irreducible representations for generators of $I4/mmm$ space group, associated with the $\mathbf{k}=(1,1,1)$ ($M$-point) and the $\mathbf{k}=(1/2,0,0), (-1/2,0,0), (0,-1/2,0), (0,1/2,0)$ wavevector star ($\Sigma$-line of symmetry) \cite{Aroyo2006}. Note that here $T$ is the time reversal operator.}
\centering 
\begin{tabular*}{0.98\textwidth}{@{\extracolsep{\fill}} c c c c} 
\hline\hline\\ 
Symm. & $mM_5^+(\delta_1,\delta_2)$ & $m\Sigma_3(\xi_1,\xi_1^*,\xi_2^*,\xi_2)$ & $\Sigma_4(\eta_1,\eta_1^*,\eta_2^*,\eta_2)$ \\ [1.5ex] 
\hline\\ 
$\left \{ 1|0,0,0\right \}$ & $\begin{pmatrix} 1 & 0 \\ 0 & 1 \end{pmatrix}$ & $\begin{pmatrix} 1 & 0 & 0 & 0 \\ 0 & 1 & 0 & 0 \\ 0 & 0 & 1 & 0 \\ 0 & 0 & 0 & 1 \end{pmatrix}$ & $\begin{pmatrix} 1 & 0 & 0 & 0 \\ 0 & 1 & 0 & 0 \\ 0 & 0 & 1 & 0 \\ 0 & 0 & 0 & 1 \end{pmatrix}$ \\ 
$\left \{ 2_z|0,0,0\right \}$ & $\begin{pmatrix} -1 & 0 \\ 0 & -1 \end{pmatrix}$ & $\begin{pmatrix} 0 & -1 & 0 & 0 \\ -1 & 0 & 0 & 0 \\ 0 & 0 & 0 & -1 \\ 0 & 0 & -1 & 0 \end{pmatrix}$ & $\begin{pmatrix} 0 & 1 & 0 & 0 \\ 1 & 0 & 0 & 0 \\ 0 & 0 & 0 & 1 \\ 0 & 0 & 1 & 0 \end{pmatrix}$ \\
$\left \{ 2_y|0,0,0\right \}$ & $\begin{pmatrix} 0 & 1 \\ 1 & 0 \end{pmatrix}$ & $\begin{pmatrix} 0 & 1 & 0 & 0 \\ 1 & 0 & 0 & 0 \\ 0 & 0 & -1 & 0 \\ 0 & 0 & 0 & -1 \end{pmatrix}$ & $\begin{pmatrix} 0 & -1 & 0 & 0 \\ -1 & 0 & 0 & 0 \\ 0 & 0 & -1 & 0 \\ 0 & 0 & 0 & -1 \end{pmatrix}$ \\
$\left \{ 4_z^+|0,0,0\right \}$ & $\begin{pmatrix} 0 & -1 \\ 1 & 0 \end{pmatrix}$ & $\begin{pmatrix} 0 & 0 & 1 & 0 \\ 0 & 0 & 0 & 1 \\ 0 & -1 & 0 & 0 \\ -1 & 0 & 0 & 0 \end{pmatrix}$ & $\begin{pmatrix} 0 & 0 & 1 & 0 \\ 0 & 0 & 0 & 1 \\ 0 & 1 & 0 & 0 \\ 1 & 0 & 0 & 0 \end{pmatrix}$ \\
$\left \{ -1|0,0,0\right \}$ & $\begin{pmatrix} 1 & 0 \\ 0 & 1 \end{pmatrix}$ & $\begin{pmatrix} 0 & 1 & 0 & 0 \\ 1 & 0 & 0 & 0 \\ 0 & 0 & 0 & 1 \\ 0 & 0 & 1 & 0 \end{pmatrix}$ & $\begin{pmatrix} 0 & 1 & 0 & 0 \\ 1 & 0 & 0 & 0 \\ 0 & 0 & 0 & 1 \\ 0 & 0 & 1 & 0 \end{pmatrix}$ \\
$\left \{ 1|1,0,0\right \}$ & $\begin{pmatrix} 1 & 0 \\ 0 & 1 \end{pmatrix}$ & $\begin{pmatrix} e^{\pi i} & 0 & 0 & 0 \\ 0 & e^{-\pi i} & 0 & 0 \\ 0 & 0 & e^{-\pi i} & 0 \\ 0 & 0 & 0 & e^{\pi i} \end{pmatrix}$ & $\begin{pmatrix} e^{\pi i} & 0 & 0 & 0 \\ 0 & e^{-\pi i} & 0 & 0 \\ 0 & 0 & e^{-\pi i} & 0 \\ 0 & 0 & 0 & e^{\pi i} \end{pmatrix}$ \\
$\left \{ 1|0,1,0\right \}$ & $\begin{pmatrix} 1 & 0 \\ 0 & 1 \end{pmatrix}$ & $\begin{pmatrix} 1 & 0 & 0 & 0 \\ 0 & 1 & 0 & 0 \\ 0 & 0 & 1 & 0 \\ 0 & 0 & 0 & 1 \end{pmatrix}$ & $\begin{pmatrix} 1 & 0 & 0 & 0 \\ 0 & 1 & 0 & 0 \\ 0 & 0 & 1 & 0 \\ 0 & 0 & 0 & 1 \end{pmatrix}$ \\
$\left \{ 1|0,0,1\right \}$ & $\begin{pmatrix} 1 & 0 \\ 0 & 1 \end{pmatrix}$ & $\begin{pmatrix} 1 & 0 & 0 & 0 \\ 0 & 1 & 0 & 0 \\ 0 & 0 & 1 & 0 \\ 0 & 0 & 0 & 1 \end{pmatrix}$ & $\begin{pmatrix} 1 & 0 & 0 & 0 \\ 0 & 1 & 0 & 0 \\ 0 & 0 & 1 & 0 \\ 0 & 0 & 0 & 1 \end{pmatrix}$ \\
$\left \{ 1|1/2,1/2,1/2\right \}$ & $\begin{pmatrix} -1 & 0 \\ 0 & -1 \end{pmatrix}$ & $\begin{pmatrix} e^{\frac{1}{2}\pi i} & 0 & 0 & 0 \\ 0 & e^{-\frac{1}{2}\pi i} & 0 & 0 \\ 0 & 0 & e^{-\frac{1}{2}\pi i} & 0 \\ 0 & 0 & 0 & e^{\frac{1}{2}\pi i} \end{pmatrix}$ & $\begin{pmatrix} e^{\frac{1}{2}\pi i} & 0 & 0 & 0 \\ 0 & e^{-\frac{1}{2}\pi i} & 0 & 0 \\ 0 & 0 & e^{-\frac{1}{2}\pi i} & 0 \\ 0 & 0 & 0 & e^{\frac{1}{2}\pi i} \end{pmatrix}$ \\
$T$ & $\begin{pmatrix} -1 & 0 \\ 0 & -1 \end{pmatrix}$ & $\begin{pmatrix} -1 & 0 & 0 & 0 \\ 0 & -1 & 0 & 0 \\ 0 & 0 & -1 & 0 \\ 0 & 0 & 0 & -1 \end{pmatrix}$ & $\begin{pmatrix} 1 & 0 & 0 & 0 \\ 0 & 1 & 0 & 0 \\ 0 & 0 & 1 & 0 \\ 0 & 0 & 0 & 1 \end{pmatrix}$ \\[1.5ex]
\hline
\hline
\end{tabular*}
\label{tab:DmitryTab}
\end{table*}

Assuming the most likely scenario, that the low temperature structural transition is a lock-in type, one can deduce the possible symmetries of the crystal structure below $T_{\rm{S2}}$. This scenario sets symmetry constraints on the transformational properties of the commensurate order parameter, impaling the same active irrep for both incommensurate and commensurate phases. As mentioned above, the modulated orthorhombic structure is associated with $\Sigma_4$. This irrep is four-dimensional with the components of the complex order parameter $(\eta_{1},\eta^{\star}_{1},\eta^{\star}_{2},\eta_{2})$. Using the matrix operators summarised in table \ref{tab:DmitryTab}, one can verify the existence of a lock-in free-energy invariant, $\eta_{1}^{4}+\eta_{1}^{\star 4}+\eta_{2}^{\star 4}+\eta_{2}^{4}$, for the commensurate value of the propagation vector $\mathbf{k}_{\rm{C}}=(1/2,0,0)$. This energy term exists only for the single value of the propagation vector and therefore its activation favours the commensurate phase. This symmetry argument further supports the lock-in mechanism for the transition at $T_{\rm{S2}}$.

In the commensurate limit of $\mathbf{k}_{\rm{C}}=(1/2,0,0)$ there are three isotropy subgroups associated with $\Sigma_4$. Depending on choice of the global phase of the modulation, one can obtain the $Pnma$, $Pmma$ and $Pmc2_1$ subgroups. They were the primary candidates in the refinement of the low-temperature diffraction data. Independent analysis based on testing of all possible isotropy subgroups of $I4/mmm$ consistent with the doubled unit cell also resulted only into two possibilities $Pnma$ and $Pmc2_1$.

\begin{figure}[!ht]
\centering
    \includegraphics[scale=0.62]{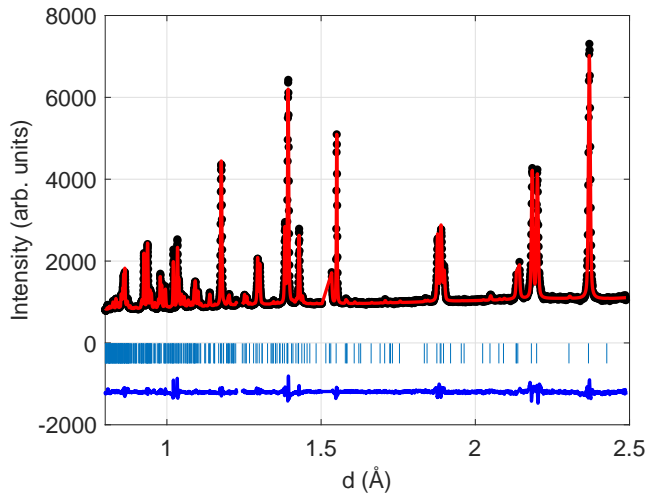}
    \caption{Data and refinement using $Pnam$ for data collected on HRPD at 20\,K as a function of d-spacing. Black circles indicate the measured data, the red line the refinement, and the blue line indicates the difference between the two (offset by -2000 units). Vertical ticks indicate the positions of Bragg reflections.}
    \label{f:pnam_refmnt}
\end{figure}

We performed refinements of all data collected on HRPD (since it had the highest resolution) at 20\,K using $Pnam$ and found generally satisfactory fits, with $R_{p}=2.51\%$ and $R_{wp}=3.33\%$. We also investigated refining the data with the $Pna2_{1}$ space group (No. 33), which is a sub-group of $Pnam$, and found a slightly improved residuals of $R_{p}=2.47\%$ and $R_{wp}=3.29\%$. However, given that the improvement in goodness-of-fit was very small and likely arises due to fewer symmetry constraints on the structure, we take the higher symmetry structure to be the solution. The final refinements using the $Pnam$ space group are shown in fig. \ref{f:pnam_refmnt}.

\begin{figure}[!ht]
\centering
    \includegraphics[scale=0.35]{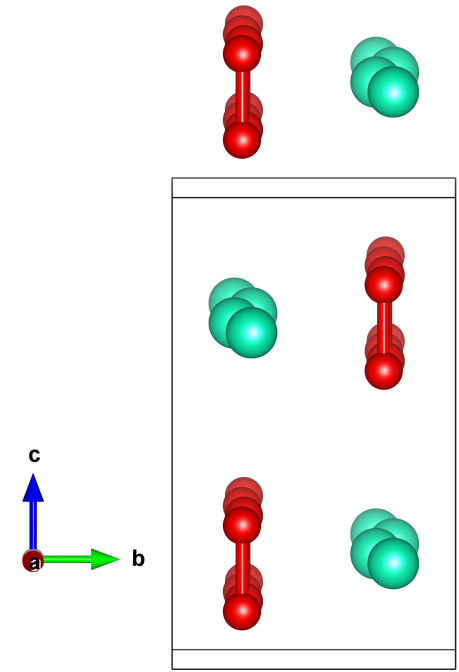}
    \caption{The refined crystal structure with the $Pnam$ space group, with oxygen molecules shown in red (spheres connected by a cylinder to indicate the molecular bond) and cesium ions shown in green (spheres only). The unit cell is indicated by the gray box. The view shown is slightly tilted from being parallel to the $a$-axis. The periodic displacements of the Cs ions are clear, with a smaller but still visible periodic displacement of the O$_{2}^{-}$ molecules also. The figure shows two unit cells along the $a$-axis.}
    \label{f:pnam_struc}
\end{figure}

\begin{table}
\caption{Crystal structure parameters for the $Pnam$ space group solution, at 20\,K measured on the HRPD instrument. The determined lattice parameters were $a = 8.7271(1) \AA$, $b = 4.39758(6) \AA$, and $c = 7.33860(8)\AA$. The unit cell volume was $V = 281.640(6) \AA^{3}$. The residuals for the fit were $R_{p}=2.51\%$ and $R_{wp}=3.33\%$.}
\centering
\begin{tabular}{ p{10mm}p{8mm}p{15mm}p{15mm}p{15mm}p{15mm}  }
 \hline \hline
 \vspace{1mm} & \vspace{1mm} & \vspace{1mm} & \vspace{1mm} & \vspace{1mm} & \vspace{1mm} \\
 Atom & Site & $x$ & $y$ & $z$ & $U_{\rm{iso}} (\AA^{2})$\\
 \vspace{1mm} & \vspace{1mm} & \vspace{1mm} & \vspace{1mm} & \vspace{1mm} & \vspace{1mm} \\
 \hline
 \vspace{1mm} & \vspace{1mm} & \vspace{1mm} & \vspace{1mm} & \vspace{1mm} & \vspace{1mm} \\
 Cs & $4c$ & 0.1207(4) & 0.7142(4) & 0.25 & 0.0022(4)\\
 O & $8d$ & 0.8747(4) & 0.2472(4) & 0.15944(8) & 0.0082(3)\\
 \vspace{1mm} & \vspace{1mm} & \vspace{1mm} & \vspace{1mm} & \vspace{1mm} & \vspace{1mm} \\
 \hline
\end{tabular}
\label{tab:Pnam_struc}
\end{table}

The details of the refinements using the $Pnam$ space group is given in table \ref{tab:Pnam_struc}, and shown graphically in fig. \ref{f:pnam_struc}. The most notable feature is that the Cs ions are shifted considerably from their ideal position of $\frac{3}{4}$ on the $b$-axis, and form a zig-zag pattern in the doubled unit cell along $a$. In the $Pnam$ structure the oxygen dumbbells are not allowed by symmetry to tilt, whereas they are in principle free to do this in the $Pna2_{1}$ structure that was investigated. In the former structural solution we found that there was a small staggered shift of the centre of mass of the oxygen dumbbells along the $b$-axis, albeit of much smaller magnitude than the shift of the Cs ions. For the $Pna2_{1}$ structure (see table \ref{tab:Pna21_struc}) this shift of the oxygen dumbbells was found to be of the same nature, i.e. there was essentially no tilt of the oxygen dumbbells even when this would be allowed by symmetry.

\begin{table}
\caption{Crystal structure parameters for the $Pna2_{1}$ space group solution, at 20\,K measured on the HRPD instrument. The determined lattice parameters were $a = 8.727(3) \AA$, $b = 4.3968(2) \AA$, $c = 7.3380(2) \AA$. The unit cell volume was $V = 281.553(3) \AA^{3}$. The residuals for the fit were $R_{p}=2.47\%$ and $R_{wp}=3.30\%$, which represent a very small improvement compared to the refinement shown in table \ref{tab:Pnam_struc} for the $Pnam$ crystal structure.}
\centering
\begin{tabular}{ p{9mm}p{8mm}p{15mm}p{15mm}p{16mm}p{15mm}  }
 \hline \hline
 \vspace{1mm} & \vspace{1mm} & \vspace{1mm} & \vspace{1mm} & \vspace{1mm} & \vspace{1mm} \\
 Atom & Site & $x$ & $y$ & $z$ & $U_{\rm{iso}} (\AA^{2})$\\
 \vspace{1mm} & \vspace{1mm} & \vspace{1mm} & \vspace{1mm} & \vspace{1mm} & \vspace{1mm} \\
 \hline
 \vspace{1mm} & \vspace{1mm} & \vspace{1mm} & \vspace{1mm} & \vspace{1mm} & \vspace{1mm} \\
 Cs & $4a$ & 0.1203(3) & 0.7816(9) & -0.0003(4) & 0.0043(4)\\
 O1 & $4a$ & 0.1277(4) & 0.7502(12) & 0.4091(14) & 0.0066(3)\\
 O2 & $4a$ & 0.6293(4) & 0.7575(8) & -0.4091(12) & 0.0091(5)\\
 \vspace{1mm} & \vspace{1mm} & \vspace{1mm} & \vspace{1mm} & \vspace{1mm} & \vspace{1mm} \\
 \hline
\end{tabular}
\label{tab:Pna21_struc}
\end{table}

\subsection{Magnetic structure}

\begin{figure}[!h]
\centering
    \includegraphics[scale=0.58]{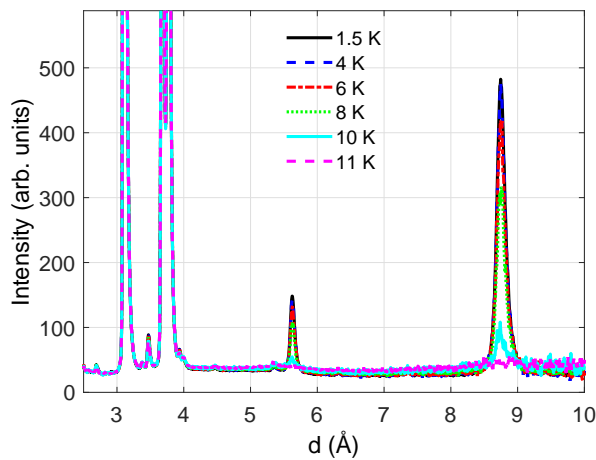}
    \caption{Overview of the data collected in the low angle bank on WISH as a function of temperature between 1.5\,K and 11\,K. Data have been converted from time-of-flight to d-spacing. The strong peaks at $\sim 8.8$\,\AA\, and $\sim 5.7$\,\AA\, correspond to the (1,0,0) and (1,0,1) Bragg reflections respectively.}
    \label{f:mag_tdep_wide}
\end{figure}

On cooling below $T \approx 10$\,K additional low $Q$ / long $d$-spacing peaks appear that are consistent with the previously supposed appearance of antiferromagnetic order. Such peaks were consistently visible in the data collected on WISH, E2, E6 and E9. As mentioned in sec. \ref{sec:methods}, only the WISH data were used for refining the magnetic structure. Of all the datasets, those from this instrument had the best signal to noise ratio and hence the most magnetic Bragg peaks were discernable, giving the greatest chance of a reliable refinement of the magnetic structure. Subsequent checks of the refinement using the E2, E6 and E9 instruments yielded good fits. An overview of the data collected as a function of temperature below 11\,K is shown in fig. \ref{f:mag_tdep_wide}.

The appearance of strong peaks at $(1,0,0)$ and $(1,0,1)$ indicates the onset of antiferromagnetic order, at $10 < T_{\rm{N}} < 11$\,K. This compares with the peaks in the inverse susceptibility and specific heat, that are taken to give the value of $T_{\rm{N}}$, of 9.6\,K \cite{Riyadi2011,Labhart,ZumstegThesis}. The small difference may be due to small differences in thermometry or a slight lag on the true sample temperature since the measurements were collected on warming.

\begin{table}[!h]
\caption{Irreducible representations of a $\mathbf{k}=(0,0,0)$ magnetic structure in the $Pnam$ space group.}
\centering
\begin{tabular}{ p{1cm}p{1.5cm}p{1.7cm}p{1.7cm}p{1.7cm} }
\hline
\vspace{0.1mm} & \vspace{0.1mm} & \vspace{0.1mm} & \vspace{0.1mm} & \vspace{0.1mm}\\
Irrep & $\mathbf{S_{k}}(1)$ & $\mathbf{S_{k}}(2)$ & $\mathbf{S_{k}}(3)$ & $\mathbf{S_{k}}(4)$ \\
\hline
\vspace{0.1mm} & \vspace{0.1mm} & \vspace{0.1mm} & \vspace{0.1mm} & \vspace{0.1mm}\\
$\Gamma_{1}(4c)$ & $(0,0,u)$ & $(0,0,u)$ & $(0,0,-u)$ & $(0,0,-u)$\\
$\Gamma_{2}(4c)$ & $(u,v,0)$ & $(-u,-v,0)$ & $(u,-v,0)$ & $(-u,v,0)$\\
$\Gamma_{3}(4c)$ & $(0,0,u)$ & $(0,0,u)$ & $(0,0,u)$ & $(0,0,u)$\\
$\Gamma_{4}(4c)$ & $(u,v,0)$ & $(-u,-v,0)$ & $(-u,v,0)$ & $(u,-v,0)$\\
$\Gamma_{5}(4c)$ & $(u,v,0)$ & $(u,v,0)$ & $(u,-v,0)$ & $(u,-v,0)$\\
$\Gamma_{6}(4c)$ & $(0,0,u)$ & $(0,0,-u)$ & $(0,0,-u)$ & $(0,0,u)$\\
$\Gamma_{7}(4c)$ & $(u,v,0)$ & $(u,v,0)$ & $(-u,v,0)$ & $(-u,v,0)$\\
$\Gamma_{8}(4c)$ & $(0,0,u)$ & $(0,0,-u)$ & $(0,0,u)$ & $(0,0,-u)$\\
\hline
\vspace{0.1mm} & \vspace{0.1mm} & \vspace{0.1mm} & \vspace{0.1mm} & \vspace{0.1mm}\\
\end{tabular}
\label{tab:irreps_pnam}
\end{table}

All possible irreducible representations of a magnetic structure with $\mathbf{k}=(0,0,0)$ in the $Pnam$ space group are given in table \ref{tab:irreps_pnam}. The general expressions of the Fourier coefficients $\mathbf{S_{k}}(j)$ were obtained from the basis functions calculated from the different representations of the O$_{2}^{-}$ units at the Wyckoff position $4c$ of the space group $Pnam$: O1 at $(x,y,\frac{1}{4})$, O2 at $(-x,-y,\frac{3}{4})$, O3 at $(\frac{1}{2}+x,\frac{1}{2}-y,\frac{1}{4})$, and O4 at $(\frac{1}{2}-x,\frac{1}{2}+y,\frac{3}{4})$. Instead of the Wyckoff position $8d$, where the individual oxygen atoms are located, we used $4c$ which defines the center of gravity of the O$_{2}^{-}$ unit.

We also considered a solution to the magnetic structure using the lower symmetry space group $Pna2_{1}$. The difference between the magnetic structures allowed in the two space groups is that $Pnam$ does not permit magnetic order with a general component of the moment in the $ab$-plane and along the $c$-axis, but one or the other. On the other hand, $Pna2_{1}$ does in principle allow magnetic moments to point along a general direction. However, we found that even refining the $c$-axis component of the magnetic moment, permitted in $Pna2_{1}$, we found the best fit yielded this component as zero within errorbar. For the following we therefore restrict our discussion just to $Pnam$.

Note that the magnetic form factor for the $s=1/2$ O$_{2}^{-}$ ions was included using the tabulated values obtained from measurements of solid oxygen \cite{Stephens_FF}. Although the size of the oxygen molecules is different (1.29\,\AA\, for solid oxygen vs 1.33\,\AA\, for CsO$_{2}$) the difference is small enough that we have reasonable confidence that the derived form factor is applicable here.

Before fitting the data we note that the strong intensity at (1,0,0) and (1,0,1) implies that the largest component of the magnetic moment will be along the $b$-axis, since the magnetic neutron diffraction cross-section is proportional to the component of the magnetic moment perpendicular to the wavevector. This means that we can immediately rule out the $\Gamma_{1}$, $\Gamma_{3}$, $\Gamma_{6}$ and $\Gamma_{8}$ irreps, since they contain moments parallel to the $c$-axis only.

We also note that for the $\Gamma_{7}$ irrep the $y$-components of the moments are coupled ferromagnetically. Since the bulk susceptibility follows a purely antiferromagnetic trend below $T_{\rm{N}}$, for the refinements this component must be fixed to zero, which immediately allows us to rule it out. This then leaves the following possible irreps: $\Gamma_{2}$, $\Gamma_{4}$, and $\Gamma_{5}$, which all have AFM arrangements of spins for the major $y$-component of the moment.

\begin{figure}[!h]
\centering
    \includegraphics[scale=0.6]{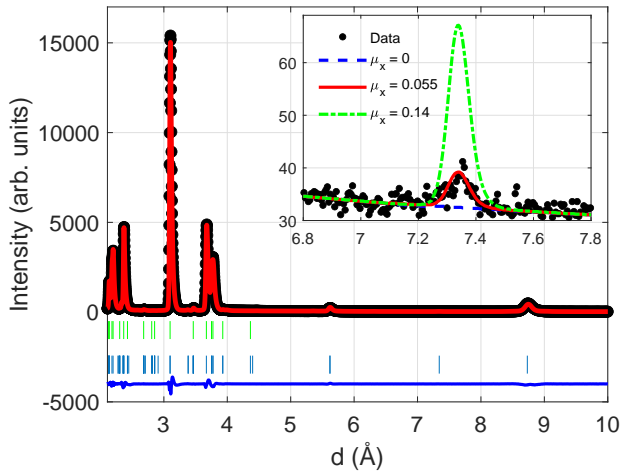}
    \caption{Data collected in the low angle bank on WISH at 1.5\,K (black points), together with the final refinement (solid red line), difference curve (solid blue line). The lower ticks (blue) are for the antiferromagnetic structure and the upper ticks (green) are for the nuclear structure. The inset shows data focused on the region where the $(0,0,1)$ peak is to be found. The lines correspond to simulations of the scattering with different values of $\mu_{x}$, with all other parameters from the refinement fixed. This shows that the intensity of the $(0,0,1)$ peak is rather sensitive to the value of $\mu_{x}$.}
    \label{f:mag_refinement_001}
\end{figure}

\begin{figure}[!h]
\centering
    \includegraphics[scale=0.4]{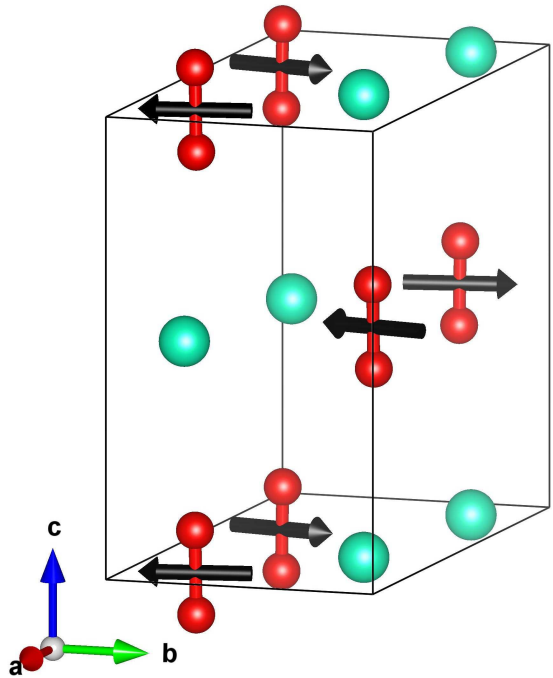}
    \caption{The refined magnetic structure of \cso, with black arrows indicating the moment direction on the O$_{2}^{-}$ dumbbells, the oxygen ions shown as red spheres, connected by a red cylinder to indicate the molecular bond, and the Cs ions are shown as green spheres.}
    \label{f:mag_struc}
\end{figure}

The presence of a very weak peak corresponding to $(0,0,1)$, that appears below $T_{\rm{N}}$ and hence must be magnetic, also allows us to narrow down our choice of possible irreps. This observation allows us to rule out $\Gamma_{5}$, for which this peak has zero intensity. Comparing $\Gamma_{2}$ and $\Gamma_{4}$, we found substantially better fits ($R_{wp}=6.36\%$) for the case of the $\Gamma_{2}$ irrep, compared to the $\Gamma_{4}$ irrep ($R_{wp}=7.75\%$). The best fit was found with $\mu_{y} = 0.512(16) \mu_{\rm{B}}$ and $\mu_{x} = 0.05(2) \mu_{\rm{B}}$, giving a total moment of $0.514 \mu_{\rm{B}}$. As illustrated in the inset of fig. \ref{f:mag_refinement_001}, the presence of the $(0,0,1)$ peak is crucial for determining the $x$-component of the magnetic moment, as the intensity is rather sensitive to the value of $\mu_{x}$. The final refinement is shown in fig. \ref{f:mag_refinement_001}, together with an inset showing the sensitivity of the $(0,0,1)$ peak's intensity to $\mu_{x}$. The final refined magnetic structure is shown in fig. \ref{f:mag_struc}. Our result implies the magnetic structure has $Pn'a'm'$ symmetry, keeping the setting of the paramagnetic $Pnam$ space group.

\subsection{Spin flop transition}

\begin{figure}[!h]
\centering
    \includegraphics[scale=0.45]{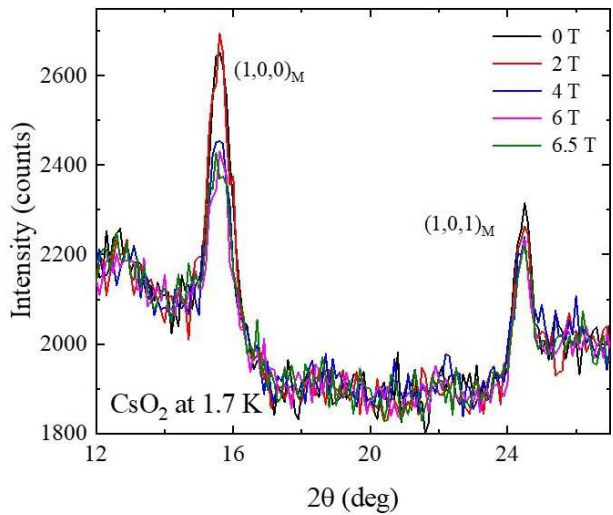}
    \caption{Data collected on E2 at $T = 1.7$\,K in applied magnetic fields between 0\,T and 6.5\,T. A sharp drop in peak intensity of the (1,0,0) and (1,0,1) magnetic Bragg reflections is observed for $2 < B < 4$\,T, most likely arising from a spin flop transition.}
    \label{f:spinflop}
\end{figure}

Data taken on the E2 diffractometer probed the effect of an applied magnetic field on the two strongest magnetic Bragg peaks, $(1,0,0)$ and $(1,0,1)$, and are shown in fig. \ref{f:spinflop}. Both peaks showed marked decreases in intensity between 2\,T and 4\,T. The strong $(1,0,0)$ peak intensity reduces in magnitude above background by approximately one third, with a somewhat smaller reduction of around $15\%$ seen for the $(1,0,1)$ peak. In the zero field phase we have established that the moments lie predominantly along the $b$-axis, and in a spin flop transition we would expect them to switch to being parallel to the $a$ or $c$ directions. Crucially, this would apply only to those (randomly oriented) crystallites in the powder sample whose $b$ axes lie mostly parallel to the applied magnetic field. Above the critical field we would therefore expect approximately one third of the crystallites to undergo a spin flop transition. The maximum change in magnetic Bragg peak intensity would therefore be $\frac{1}{3}$, corresponding to the signal from the one third of crystallites involved in the spin flop going to zero. Because the $(1,0,0)$ peak reduces by approximately this amount we can surmise that the spin flop involves spins reorienting from the $b$-axis to the $a$-axis. If the spins were to flop towards $c$, and assuming no change in the size of the ordered moment, then the component of magnetization perpendicular to $\mathbf{Q}$ would be unchanged, and hence no change in the intensity of the $(1,0,0)$ Bragg peak would be seen, at variance with our observations. On the other hand, a spin flop towards $a$ would result in the component of magnetization perpendicular to $\mathbf{Q}$ going to zero. We can check this result by looking at the intensity of the $(1,0,1)$ magnetic Bragg peak as well. Spins reorienting from $b$ to $a$ would result in a reduction in Bragg peak intensity of $\sim 40\%$ for the one third of crystallites involved, giving rise to an overall reduction in intensity of $\sim 13\%$, which is broadly consistent with what we observe.

\section{Discussion and Conclusion}\label{sec:discussion}

We have determined the low temperature crystal structure of CsO$_{2}$, finding it to be described by the orthorhombic space group $Pnam$ (or equally well by the lower symmetry orthorhombic space group $Pna2_{1}$) in which the $a$ axis is approximately double the length of the $b$ axis. Earlier work had indicated a lowering of symmetry on cooling from room temperature, for example via the appearance of extra peaks in the Raman spectrum that could not be accounted for in a high symmetry setting \cite{Riyadi2012} or due to the observation of the weak $(\frac{1}{2},0,0)$ peaks (indexed in the high temperature tetragonal unit cell) at low temperature in x-ray diffraction measurements \cite{Ziegler1976}. Furthermore, DFT calculations have predicted an enlarged (doubled) unit cell compared to the high temperature tetragonal phase \cite{Riyadi2012}. However, we note neither that the Raman nor the x-ray data were used to determine the detailed crystal structure as we have here. In addition, the DFT calculations were correct in predicting a doubled unit cell, but suggested that the doubling occurs along more than one crystallographic axis, which we did not observe.

Concerning the details of the low temperature crystal structure we determined, we see that the doubling of the unit cell along $a$ is predominantly due to a zig-zag displacement (`puckering') of the Cs ions, with a smaller zig-zag displacement of the centers of mass of the O$_{2}^{-}$ dumbbells. The staggered displacement of Cs ions was foreseen by DFT \cite{Riyadi2012}. On the other hand, the same DFT also predicted a tilting relative to the $c$-axis of the O$_{2}^{-}$ dumbbells, a scenario with which our data are inconsistent.

Looking at the refined low temperature commensurate structure we can compare it to the incommensurate structure visible for $80 \simeq T \simeq 200$\,K by looking at the amplitude of the displacement of the Cs ions and O$_{2}^{-}$ dumbbells respectively. The amplitude of the sinusoidal modulation of Cs ion positions in the incommensurate phase was 0.0564(9) lattice units, whereas the amplitude of the modulation, given by the deviation of the Cs ion position along the $b$-axis relative to the ideal coordinate of $y=0.75$ in the commensurate phase, is 0.0358(1) lattice units. These values are rather comparable. The amplitude of the O$_{2}^{-}$ dumbbell displacement was 0.0090(7) lattice units in the incommensurate phase, whereas in the commensurate phase it is 0.0028(4) lattice, which is somewhat different.

We see in fig. \ref{f:Tdep_latt} that cooling below the transition from a tetragonal phase at $T_{\rm{S1}}=192$\,K the $a$ lattice parameter of the average structure shrinks significantly more rapidly down to $T_{\rm{S2}}=71.5$\,K, reducing by $\sim 1.2\%$ compared to the $b$ lattice parameter which reduces by just $0.34\%$ and the $c$ lattice parameter which barely changes ($< 0.1\%$ change). At $T_{\rm{S2}}$ the $a$ lattice parameter (when indexed using the $Immm$ space group) suddenly increases. A possible explanation for this is that the puckering of the Cs ions in the incommensurate structure at comparatively high temperature, when the oxygen molecules' orientation is fluctuating significantly, results in strain on the lattice due to under-bonding. As the fluctuations decrease on cooling eventually the Cs ions and oxygen molecules can bond, leading to a lock-in to the commensurate structure, a decrease in lattice strain and a corresponding slight increase in $a$.

It was already noted that a staggered structure would likely have an impact on the strength of the superexchange interaction between the $s=1/2$ units, due to differing orbital overlap between the oxygens and the cesium ions \cite{Klanjsek2015}. NMR data modelled under the assumption of staggered tilts of the O$_{2}^{-}$ dumbbells indicated the formation of antiferromagnetic 1d spin chains in \cso. The temperature dependence of the magnetic exchange parameters determined from the NMR was understood in relation to the assumed staggered dumbbell tilts together with a model based changing orbital overlap as a result of librations of the dumbbells. Our findings show that in fact the dumbbells are hardly tilted at all, rather the cesium ion positions are staggered. One might anticipate that the overall effect could be similar, at least as far as superexchange is concerned. It would be interesting to re-model the NMR data based on the crystal structure we have determined.

We note that there is some discussion in the literature of orbital ordering in superoxides \cite{Klanjsek2015,Knaflic2015,Nandy2010,Kim2010,Kim2014}, which has an impact on the aforementioned spin chain formation. Orbital order is frequently accompanied by structural distortion, as in the cooperative Jahn-Teller effect \cite{Kanamori-coopJT-1960}. In the tetragonal $I4/mmm$ structure of CsO$_2$, one can take the $2b(0,0,1/2)$ Wyckoff position with the $4/mmm$ site symmetry as the place where the O$_{2}^{-}$ molecules reside. The degenerate $\pi^*$ molecular orbitals of O$_{2}^{-}$ form basis of two-dimensional $E_u$ site symmetry irrep, which has non-zero subduction frequency in the $\Sigma_4$ space group irrep. This implies that local distortions that belong to the site symmetry irrep $E_u$ can induce global distortions that belong to the space group irrep $\Sigma_4$. In other words, the structural distortions obtained in the present work are compatible with the orbital ordering scenario. The ground state orthorhombic structure is consistent with the orbital pattern proposed by Riyadi et al. \cite{Riyadi2012} although no additional doubling along the $b-$ and $c-$axes suggested by the authors, due to tilting of the oxygen dimers, was detected in our diffraction data. The incommensurate phase implies partial occupancies of the $\pi_x^*$ and $\pi_y^*$ orbitals, modulated upon propagation through the crystal (i.e. an orbital density wave). A similar phenomenon takes place in some perovskite manganites containing octahedrally coordinated Mn$^{3+}$ cations with electronic degeneracy. The modulated orbital states in these systems vary from achiral density waves \cite{Perks2012} to chiral orbital helices \cite{Khalyavin2020}.

Considering now the magnetic structure, we have confirmed that, as long anticipated, \cso is an antiferromagnet. We note that until now the details of the magnetic structure had remained unknown, with the only proposal being that of ferromagnetism within the $ab$-plane with an antiferromagnetic stacking between planes \cite{Labhart}, which was formed on the basis of analogy with KO$_{2}$ \cite{Smith1966} rather than a direct measurement. So, our data clarify this point at last, revealing ferromagnetism in the $bc$-plane and antiferromagnetism along $a$. The total magnetic moment determined, $0.514\mu_{\rm{B}}$, is reduced compared to the ideal spin only value, possibly due to fluctuations and/or reduced dimensionality \cite{Klanjsek2015}.

The presence of a small component of the magnetic moment along the $a$ axis, as well as the larger component along $b$, is significant. These orthogonal spin components are transformed by the same irrep of the paramagnetic $Pnma$ space group (table \ref{tab:irreps_pnam}), implying bi-linear coupling between them. This is a typical case of antisymmetric Dzyaloshinskii-Moriya exchange underpinning the coupling at microscopic level \cite{DZYALOSHINSKY}. Let us denote the corresponding magnetic order parameters as $\delta$ and $\xi$, respectively. The bi-linear invariant, $\delta \xi$, in principle also involves single ion terms, however these are not relevant here due to the $S=1/2$ nature of the interacting spins. The $\delta$ and $\xi$ order parameters are transformed by distinct irreps ($mM_5^+$ and $m\Sigma_3$) of the tetragonal $I4/mmm$ space group (table \ref{tab:DmitryTab}), indicating that they would be decoupled if no structural distortion was present. Decomposition of the experimentally determined ground state crystal structure of CsO$_2$ in respect of symmetrised displacive modes of the tetragonal $I4/mmm$ reveals the presence of $\Gamma^+_1 (k=0), \Gamma^+_2 (k=0), M_5^-(k=1,1,1)$ and $\Sigma_4 (k=1/2,0,0)$ modes. The latter has the largest amplitude (of 0.25$\AA$), as expected for the primary order parameter. Further analysis of the allowed free-energy terms reveals that the $\Sigma_4(\eta_1,\eta_1^*,\eta_2^*,\eta_2)$ order parameter forms a trilinear invariant with $mM_5^+(\delta_1,\delta_2)$ and $m\Sigma_3(\xi_1,\xi_1^*,\xi_2^*,\xi_2)$:
\begin{equation}
\begin{split}
\delta_1\eta_1\xi_1^*+\delta_1\eta_1^*\xi_1-\delta_1\eta_2\xi_2^*-\delta_1\eta_2^*\xi_2\\
-\delta_2\eta_1\xi_1^*-\delta_2\eta_1^*\xi_1-\delta_2\eta_2\xi_2^*-\delta_2\eta_2^*\xi_2 \nonumber
\end{split}
\end{equation}
indicating that this structural distortion is responsible for the coupling of the orthogonal magnetic modes in the ground state orthorhombic structure. The term is reduced down to $\delta_1\eta_1\xi_1^*+\delta_1\eta_1^*\xi_1-\delta_2\eta_1\xi_1^*-\delta_2\eta_1^*\xi_1$ for the case of single arm of the $\Sigma$-star, as observed experimentally. The coupling is fully optimised when the $mM_5^+(\delta_1,\delta_2)$ order parameter takes the $\delta_1=-\delta_2$ direction $(\delta_1,-\delta_2)$. This direction represents the symmetry of the experimentally observed secondary antiferromagnetic mode with the components of the spins along the orthorhombic $a$-axis.

In conclusion, we have used neutron diffraction to perform a comprehensive study of the crystal and magnetic structure of \cso. We find that an incommensurate crystal structure appears seemingly simultaneously with a transition from a tetragonal to an orthorhombic structure at 192\,K. This incommensurate structure, which is modulated along $a$, is composed of displacements of the cesium ions along the $b$-axis and of smaller displacements of the O$_{2}^{-}$ dimers' entire centre of mass. On cooling further below $72$\,K the modulated structure locks into a commensurate wavevector of $(\frac{1}{2},0,0)$, doubling the unit cell compared to the previously supposed orthorhombic crystal structure. Hints of a structure bearing some similarity to this had previously been suggested from DFT calculations. However, in both calculations and in other superoxides it was found that the O$_{2}^{-}$ dimers tilt with respect to the $c$-axis, which is different to what is seen here. Magnetic order sets in below $\sim 10$\,K and as previously supposed this order is antiferromagnetic. The spins of the O$_{2}^{-}$ dimers modulate along the $a$-axis and point predominantly along $b$, albeit with a small component along $a$. Thus, CsO$_2$ is an interesting example of a structure-properties relationship. The large structural distortion, associated with the staggered cesium and oxygen displacements, activates antisymmetric exchange and couples the orthogonal magnetic modes directly observed in the neutron diffraction experiment. For applied magnetic field $2 < B <4$\,T we see changes in magnetic Bragg peak intensity that are consistent with a spin flop transition in which the magnetic moments reorient to point along the $a$-axis.

Note: just before the completion of this manuscript we became aware of another work, ref. \onlinecite{nakano2023}, in which neutron powder diffraction was used to examine the magnetic structure of CsO$_{2}$. The authors also discuss the structural aspects of the order. The main difference to the present work compared to ref. \onlinecite{nakano2023} is in how the structural distortions are combined with the magnetic order. Taking the tetragonal $I4/mmm$ space group as the common parent symmetry, the suggested structural distortions and the magnetic order double the $a$- and $b$- axes, respectively. In the ground state found in our work, both the primary structural distortions and the spin ordering double the same tetragonal axis (the $a$-axis in our setting). In addition, the measurements presented in ref. \onlinecite{nakano2023} were insensitive to the small $a$-axis component of the magnetic moment that we found.

\begin{acknowledgements}
We are grateful to Bella Lake and Ryota Ono for insightful discussions. Experiments at the ISIS Neutron and Muon Source were supported by an Xpress beamtime allocation $\rm{RB}1990392$ from the Science and Technology Facilities Council. Raw data are available online \cite{hrpd_data}.
\end{acknowledgements}

\bibliography{CsO2_refs}

\appendix

\end{document}